\documentclass{article}
\usepackage{amsmath,graphicx,mlspconf}

\usepackage{booktabs}
\usepackage{float}
\usepackage{algorithm}
\usepackage{algorithmic}
\usepackage{newfloat}
\usepackage{listings}
\usepackage{amsmath}
\usepackage{amssymb}
\usepackage{fdsymbol}
\floatstyle{ruled}
\newfloat{listing}{tb}{lst}{}
\floatname{listing}{Listing}

\usepackage{enumitem} 
\usepackage[usenames,dvipsnames]{xcolor} 
\usepackage{soul, color}
\definecolor{xpink}{rgb}{1.0, 0.15, 0.99}
\definecolor{xgreen}{rgb}{0.33, 0.98, 0.01}
\definecolor{xorange}{rgb}{0.98, 0.65,0.01}
\definecolor{xcyan}{rgb}{0.54, 0.98,1.0}
\definecolor{lightgrey}{rgb}{0.92, 0.96, 1.0}  
\definecolor{mypink}{rgb}{1, 0.9, 1} 
\DeclareRobustCommand{\hlgray}[1]{{\sethlcolor{lightgrey}\hl{\textbf{#1}}}}
\DeclareRobustCommand{\hlpink}[1]{{\sethlcolor{mypink}\hl{\textbf{#1}}}}
\newcommand\hlc[1]{%
  \bgroup
  \hskip0pt\color{red!80!black}%
  #1%
  \egroup
}
\usepackage{blindtext}
\usepackage{stfloats}
\usepackage{supertabular}
\usepackage{xcolor}
\usepackage{pifont}
\usepackage{pdfcol}
\definecolor{dgray}{rgb}{0.4, 0.4, 0.4}

\usepackage{tabularx}
\usepackage{multirow}
\usepackage{booktabs}
\usepackage{makecell}
\usepackage{url} 
\usepackage{scalerel}
\usepackage{adjustbox}

\def\decibel{\scalerel{$d$\kern-1.4pt}{\clipbox{2pt 0pt 0pt 0pt}{B}}}
\usepackage{tikz}
\newcommand*{\priority}[1]{\begin{tikzpicture}[scale=0.15]%
    \draw (0,0) circle (1);
    \ifthenelse{#1>0}{\fill[fill opacity=0.5,fill=blue] (0,0) -- (90:1) arc (90:90-#1*3.6:1) -- cycle;}{}
    \end{tikzpicture}}
\usepackage{pdfpages} 
\usepackage{hyperref}
\hypersetup{hidelinks,}

\toappear{2024 IEEE International Workshop on Machine Learning for Signal Processing, Sept.\ 22--25, 2024, London, UK}

\title{YourMT3+: Multi-instrument Music Transcription with Enhanced Transformer Architectures and Cross-dataset Stem Augmentation}

\name{%
   Sungkyun Chang$^{\star}$%
   \qquad Emmanouil Benetos$^{\star}$%
   \qquad Holger Kirchhoff$^{~\dagger}$
   \qquad Simon Dixon$^{\star}$ \thanks{This research utilized Queen Mary's Andrena HPC facility supported by QMUL Research-IT, and the AI Industrial Convergence Cluster supported by the Ministry of Science and ICT of Korea, and Gwangju Metropolitan City. EB is supported by RAEng/Leverhulme Trust Research Fellowship LTRF2223-19-106.}%
}
\address{%
    $^{\star}$ Centre for Digital Music, Queen Mary University of London\ \ \ %
    $^{\dagger}$ Huawei%
}

\begin{document}
\ninept

\maketitle

\begin{abstract}
Multi-instrument music transcription aims to convert polyphonic music recordings into musical scores assigned to each instrument. This task is challenging for modeling as it requires simultaneously identifying multiple instruments and transcribing their pitch and precise timing, and the lack of fully annotated data adds to the training difficulties. This paper introduces \texttt{YourMT3+}, a suite of models for enhanced multi-instrument music transcription based on the recent language token decoding approach of MT3. We enhance its encoder by adopting a hierarchical attention transformer in the time-frequency domain and integrating a mixture of experts. To address data limitations, we introduce a new multi-channel decoding method for training with incomplete annotations and propose intra- and cross-stem augmentation for dataset mixing. Our experiments demonstrate direct vocal transcription capabilities, eliminating the need for voice separation pre-processors. Benchmarks across ten public datasets show our models' competitiveness with, or superiority to, existing transcription models. Further testing on pop music recordings highlights the limitations of current models. Fully reproducible code and datasets are available with demos at \url{https://github.com/mimbres/YourMT3}.

\end{abstract}

\section{Introduction}
\label{sec:intro}
Automatic music transcription (AMT)~\cite{benetos2018automatic} is a fundamental task in music information retrieval where the goal is to transform music audio input into a sequence of musical notes, with each note possessing properties such as onset, offset, pitch, and sometimes velocity. The output is typically presented in the form of MIDI or piano-roll notation. The significance of AMT extends to a wide range of applications, including interactive music systems~\cite{rowe2004machine}, automatic accompaniment generation~\cite{percival2015song2quartet}, and music performance assessment.

The key challenge of this research is multi-instrument AMT: identification and transcription of various instruments with vocals from music recordings. 
Recently, there has been notable progress in this field: MT3~\cite{mt3} utilized a MIDI-like decoding transformer, while PerceiverTF~\cite{perceivertf} employed a spectral attention transformer that generates conventional piano-roll. Unfortunately, the absence of fully reproducible code for these models has been a significant limitation for replication and further research. Our replication of MT3, trainable from scratch, is dubbed as \texttt{YourMT3}~\cite{chang2022yourmt3}.
Based on this, we  propose \texttt{YourMT3+}, a hybrid architecture that incorporates advanced architectures and training methods for further enhancements. \texttt{YourMT3+} and its variants differ from prior work~\cite{mt3, perceivertf} in the following key aspects:
\begin{itemize}[leftmargin=*]
\item \textbf{Enhanced Encoder}: PerceiverTF~\cite{perceivertf}, which generated piano-rolls, is now trained with the MT3 framework to generate note event tokens. We replaced MT3's encoder with PerceiverTF featuring spectral cross attention (SCA). Additionally, replacing its feedforward network (FFN) with a mixture of experts (MoE)~\cite{jiang2024mixtral}, denoted as \texttt{YPTF.MoE}, demonstrates promising results.

\item \textbf{Multi-channel Decoder}: In addition to \textit{General MIDI} tokens, singing transcription tokens have been further defined. We introduce a multi-channel decoder that replaces MT3's single-channel decoder~\cite{mt3}. This enables task-query based training and the use of partially annotated data, improving performance. 
\item \textbf{Augmentation}: The proposed online data augmentation framework incorporates intra-stem and cross-stem mixing across datasets and pitch-shifting. In particular, \textit{cross-stem augmentation} allows for transcribing singing with other instruments without the need for a voice separation front-end. 
\item \textbf{Evaluation}: Our models were extensively validated on various multi-instrument and single-instrument datasets. One of the main applications of multi-instrument AMT can be transcribing pop music. We provide refined annotations for the existing pop music dataset~\cite{rwc}, presenting the first study to investigate multi-instrument AMT performance on commercial pop music.
\end{itemize}

\section{Relation to Prior Work}
\label{sec:related_works}
While substantial research exists in AMT, multi-instrument transcription has recently seen significant developments. The field often faces challenges due to the scarcity of fully annotated datasets for all instruments, making it \textit{low-resourced}. Strategies such as multi-task learning \cite{mt3, wu2020multi}, unsupervised learning methods \cite{reconvat} and iterative re-alignment techniques \cite{musicnet_em} have offered partial remedies, with most models producing piano-roll outputs at the frame level.

Compared to the conventional AMT models based on onsets and frames \cite{onf}, MT3 \cite{mt3} is a sequence-to-sequence model that mainly distinguished itself in decoding outputs. It decodes a note-level representation similar to language tokens derived from MIDI, deviating from the traditional frame-level piano-rolls. In Section~\ref{sec:output_token}, we discuss the advantages of using these output tokens in \texttt{YourMT3}. 


The transcription of singing within multi-instrument AMT remains largely unexplored, despite potential overlaps with source separation \cite{spleeter} and melody extraction \cite{hsieh2019streamlined}. \textit{PerceiverTF}~\cite{perceivertf}, a model with piano-roll output, has significantly advanced the transcription of multiple instruments and vocals by introducing spectral cross-attention (SCA) and stem dataset mixing. We propose an augmentation method, denoted by a plus (+) sign, that formalizes the earlier stem mixing approach~\cite{perceivertf} within an online multi-dataset pipeline.



\begin{figure*}[ht]
\centering
\includegraphics[width=0.94\textwidth]{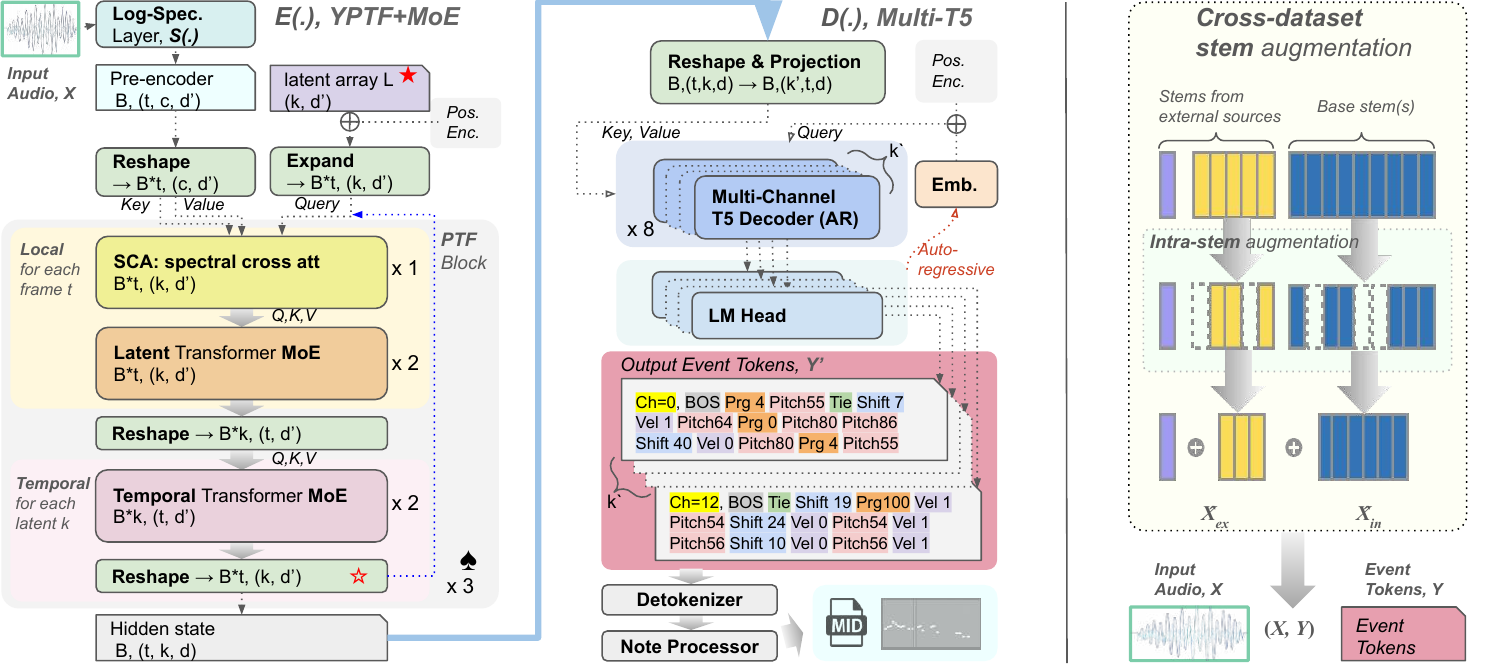} 
\caption{Overview of \textit{YourMT3+}. (left) Our encoder $E(\cdot)$ takes as input a log mel spectrogram $S$ derived from audio $X$. (center) An auto-regressive decoder $D(\cdot)$ with the language model (LM) head is conditioned by $E(S)$, and output event tokens $Y'$. (right) Cross-dataset stem augmentation, described in Section~\ref{sec:augmentation}.}  
\label{fig_overview}
\end{figure*}

\section{Model}
In the \texttt{YourMT3+} taxonomy, \texttt{YMT3} models match MT3's~\cite{mt3} architecture and training. \texttt{YPTF+Single} models use PerceiverTF (PTF) encoder with MT3's single-channel decoder and stem augmentation (\texttt{+}). Our empirical finding demonstrates that PTF's hierarchical attention with instrument-group sub-task queries enhances multi-instrument AMT in complex mixtures. \texttt{YPTF.MoE} replaces the encoder's FFN with mixture of experts (MoE), enabling task-specific encodings in multi-dataset training. These models efficiently process MIDI tokens instead of piano-roll. Our multi-channel decoder assigns instrument groups per channel and masks loss for unannotated instruments, allowing training with incomplete labels. The final \texttt{YPTF.MoE+Multi} model integrates all these features.

\label{sec:model}
The left panel of Figure~\ref{fig_overview} provides a detailed overview of our final extended model, \texttt{YPTF.MoE+Multi}. The subsequent subsections will detail the components of our model variants, including the audio input, encoder, decoder, and output tokens.

\subsection{Input}
\label{sec:input}
In Figure~\ref{fig_overview}, \(X\) represents a 2.048-second audio segment. In \texttt{YMT3}, \(X\) is transformed into a \textit{log-magnitude mel-spectrogram} \(S \in \mathbb{R}^{t \times f}\) with 256 time steps and 512 mel-frequency bins. In \texttt{YPTF}, \(X\) is initially transformed into a \textit{log-magnitude spectrogram} with 110 time steps and 1,024 frequency bins. Subsequently, a convolutional feature \(S_\mathit{conv}\) is produced by 2D ResNet pre-encoder~\cite{perceivertf}, resulting in \(S_\mathit{conv} \in \mathbb{R}^{t \times c \times f'}\), where both \(c\) and \(f'\) are set to 128. The multi-resolution input of \texttt{YPTF} mirrors \textit{PerceiverTF}~\cite{perceivertf}, including an additional channel dimension \(C\), and differs from PerceiverTF only in the input length, using 2.048 seconds instead of 6 seconds.

\subsection{Encoder}
\label{sec:encoder}
The encoder $E(\cdot)$ takes $S$ as an input, where the last dimension of $S$ typically matches the encoder's hidden dimension $d$. Our baseline encoder of \texttt{YMT3} is based on the T5-small v1.1~\cite{raffel2020exploring} encoder composed of 8 standard transformer blocks with 6-head self-attention and gated FFNs. The proposed \texttt{YPTF} replaces the encoder with PerceiverTF (PTF)~\cite{perceivertf} blocks as depicted in Figure~\ref{fig_overview} (left). 

\noindent\textbf{PTF block}: Each PTF block in our model comprises local and temporal transformer sub-blocks. The local transformer first employs spectral cross attention (SCA), derived from Perceiver~\cite{perceiver}, using a learnable latent array \(L \in \mathbb{R}^{k\times d'}\) and \(S_\mathit{conv}\) as inputs. Here, \(k\) is typically set to twice the number of target instrument groups, where \(k < c\) and specifically \(k=26\) for 12 instruments plus singing, with each pair of latents serving as a \textit{query} for the corresponding instrument groups. The latent and temporal transformer sub-blocks, featuring 8-head self-attention, FFNs and residual connections for queries, differ functionally: the former processes spectral information independently of time \(t\), by attending to \(k\) and \(c\), whereas the latter handles only temporal information relevant to \(t\) and \(d\), independent of \(k\). Overall, the PTF block (\(\spadesuit\), Figure \ref{fig_overview}) performs three iterations. Initially, \(\medstar\) acts as the \textit{query} in SCA during the first iteration. In the second and third iterations, \(\medwhitestar\) serves as the \textit{query}.

\noindent\textbf{MoE}: \texttt{YPTF.MoE} models replace FFNs in latent and temporal transformer blocks with MoE layers~\cite{jiang2024mixtral}, routing attention to two of eight experts. 
Using two experts gave better results than one or four; see Supplemental B.5. In our experiments, MoE increased the model complexity by about 5\% while improving performance across various datasets. Unlike PerceiverTF, we use RoPE~\cite{roformer} in every sub-block of the encoder to integrate positional information through rotation matrices, replacing trainable position embedding (PE), and pre-LayerNorm with pre-RMSNorm.~However,~these~modifications only offered minor benefits in memory and computation without significantly impacting performance.

\subsection{Output Tokens}
\label{sec:output_token}
The center panel of Figure~\ref{fig_overview} shows the output sequence \(Y'\) with a maximum \(N\) time steps, and the tokens representing MIDI-like events are listed in Supplemental F. As noted in Section~\ref{sec:main_result}, models trained with more fine-grained vocabulary consistently perform better. Therefore, we use \texttt{MT3\_FULL\_PLUS} for training and \texttt{MT3\_MIDI\_PLUS} only for comparison tests with previous work. Following the note sequence structure in MT3~\cite{mt3}, we made two modifications to the MT3 tokens: (a) unused velocity tokens, except 0 and 1, were removed, and (b) programs 100 and 101 were reserved for singing voice (melody) and singing voice (chorus), respectively.

Compared to traditional piano-rolls~\cite{onf, thickstun, reconvat, wu2020multi, perceivertf}, MIDI-like tokens~\cite{mt3} offer several advantages: they are more memory-efficient by representing note onset, shift, and offset with tokens rather than hundreds of frames; they simplify multi-instrument data handling by expanding the program vocabulary without significant memory increase, whereas piano-rolls need large separate matrices for each instrument; and they explicitly represent linked note onsets and offsets, avoiding extra post-processing required for piano-rolls.

\subsection{Decoder}
\label{sec:decoder}
We use an auto-regressive decoder \(D(\cdot)\), conditioned on the encoder's last hidden state, to generate note sequences. The baseline decoder, based on T5-small v1.1 and denoted as \texttt{Single}, produces a single sequence with events from multiple instruments.

When annotations are available for only one or some instruments in the audio, we need to mask the loss for unannotated instruments. The \texttt{Single} decoder's output blends multiple programs, making it hard to mask specific instruments due to token dependencies. To address this, we propose a \texttt{Multi} decoder. It can provide separately maskable supervision for each latent \( L \) of the PTF encoder, allocated into channels for each program group.

In our implementation, the PTF encoder's output hidden states are grouped by allocating two latents per channel---with group-linear projection, $k = 26$ latents result in $k' = 13$ projected channels. The \texttt{Multi} decoder then independently decodes each of the $k'$ inputs, producing $k'$ sequences for each program using parallel decoders with shared parameters. We set the maximum sequence length to \(N_{\text{single}} = 1024\) (as in MT3~\cite{mt3}) and \(N_{\text{multi}} = 256\). Potential truncation loss is discussed further in Supplemental B.6.


\section{Data Augmentation}
\label{sec:augmentation}
This section describes an augmentation method for training with multiple datasets. Our strategy is to maximize the diversity of the training examples by randomly mixing selected stems from across multiple datasets. \textit{Intra-stem augmentation} described in Section 4.1 involves selectively muting stems within a multi-track recording to generate several variations, as demonstrated with MT3~\cite{mt3} and the Slakh dataset. The concept of \textit{cross-dataset stem augmentation}, as discussed in Section 4.2, draws inspiration from PerceiverTF~\cite{perceivertf}. It aims to create a new mixture of stems from multiple datasets. Additionally, we employ pitch-shifting as described in Section 4.3.

\subsection{Intra-stem Augmentation}
This refers to the process of randomly dropping instruments from a segment containing multiple stems. 
From any dataset we sample \(X\), a 2.048-second segment starting from a random point.  Assuming that \(X\) is composed of \(N\) stems denoted \(x_1, x_2, \ldots, x_N\), we define a set \(\hat{X}_{\text{in}}\) of randomly selected or dropped stems as:
\begin{equation}
\hat{X}_{\text{in}} = \{x_i : x_i \in X, \text{ with } x_i \sim \text{Bernoulli}(p)\} 
    \label{eq:intra_aug}
\end{equation}
with $i \in \{1,2,...,N\}$ where $N > 1$. 
Here, \(p\)=$0.7$ by default, is the probability of each stem being selected. Each \(x_i\) is chosen with \(p\), creating \(\hat{X}_{\text{in}}\) with various combinations of stems from \(X\). A larger \(p\) increases active stems and task difficulty. The sweet spot was between 0.6 and 0.8, increasing with model size and training time.

\begin{algorithm}[!t]
\small
\caption{Cross-dataset Stem Augmentation}
\begin{algorithmic}[1]
\REQUIRE $X$, $U$, $L$, $J$, $\Psi$, $\tau$, $p$
\COMMENT{ \par
{\footnotesize $X$: A segment $X \in U$, with stems $x \in X$. \par
$U$: Cached segment batches from various datasets. \par
$L$: Maximum length of sequence. 1,024 by default. \par
$J$: Maximum number of iterations w.r.t $j$. 5 by default. \par
$\Psi$: Stem mixing policy. \par
$\tau$: Exponential decay parameter. 0.3 by default. \par
$p$: Probability for intra stem selection. 0.7 by default.}
}
\STATE $\hat{X}_{\text{in}} \leftarrow {x_i : x_i \in X, \text{selected with } x_i \sim \text{Bernoulli}(p)}$
\STATE $\hat{X}_{\text{ex}} \leftarrow \emptyset$
\STATE $j \leftarrow 0$
\WHILE{$r \sim \textit{Uniform}(0,1) < e^{-\tau j} \text{ and } |\hat{X}_{\text{ex}}| < L$  \text{ and } $j < J$}
\STATE $X' \leftarrow$ a randomly sampled segment from $U \setminus X$
\STATE $X' \leftarrow \text{Filter}(X'; \Psi)$ \textcolor{ForestGreen}{\footnotesize // retain stems meeting criteria}
\STATE
\IF{$X' \neq \emptyset$}
\STATE $\hat{X}_{\text{ex}} \leftarrow \hat{X}_{\text{ex}} \cup {X'}$ \textcolor{ForestGreen}{\footnotesize // add stems}
\STATE $j \leftarrow j + 1$
\ENDIF
\ENDWHILE
\STATE $\hat{X} \leftarrow \hat{X}_{\text{in}} \cup \hat{X}_{\text{ex}}$
\STATE $\text{Mix}(\hat{X})$ \textcolor{ForestGreen}{\footnotesize // apply stem mixing} 
\end{algorithmic}
\label{alg:cross_aug}
\end{algorithm}

\subsection{Cross-dataset Stem Augmentation}

\noindent\textbf{Procedure: }
In Algorithm~\ref{alg:cross_aug}, we designate $U$ as a collection of cached segment batches across diverse datasets, with its size required to be at least equal to the batch size and preferably larger, if permitted by memory constraints. The base segment $X$ is a sampled segment from $U$, and the elements of $X$ are stems denoted by $x$. Here, $x$ signifies a stem ID, including related token and audio information.   

\textit{Intra-stem augmentation} is first applied to $X$ as in Equation~\ref{eq:intra_aug}, yielding a processed base segment $\hat{X}_{\text{in}}$. 
Next, we enter a loop to mix the base stems of $\hat{X}_{\text{in}}$ with the stems coming from other segments. $U \setminus X$ represents the set of all segments in $U$ excluding $X$. Each iteration begins by randomly sampling a segment $X'$ from $U \setminus X$. Stems in $X'$ that do not satisfy policy $\Psi$ (detailed in Supplemental D.2) are then filtered out. Subsequently, $\hat{X}_{\text{ex}}$ is updated by merging $X'$. This loop persists until at least one stopping criterion described in the following subsection is satisfied. Once the aggregation is complete, the Mix($\cdot$) function executes the actual mixing of tokens and audio content in a batch-wise manner.

\noindent\textbf{Stopping criteria}
\label{tab:stopping_criteria}
In Line 4 of Algorithm~\ref{alg:cross_aug}, three criteria are established to stop the iterative mixing among stems. The first criterion is an exponential decay $S(j)$ that serves as the survival function defined as
\begin{math}
    S(j) = e^{-\tau j},
    \label{eq:exp_decay}
\end{math}
where $\tau$ controls the surviving curve with respect to $j$-th iteration. 
The second criterion restricts \(\hat{X}_{\text{ex}}\) to a length \(L\), measured as sequence length post-tokenisation. The last  criterion, \(j > J\) with \(J=5\) allows mixing up to 5 segments per base segment.


\subsection{Pitch-shifting}
We apply GPU-based phase vocoder pitch-shifting adapted from \texttt{TorchAudio}\footnote{\url{https://pytorch.org/audio}} after cross-dataset stem augmentation, using default settings except nFFT=512 for time-stretching. Batch elements are randomly assigned to five groups, each shifted by -2, -1, 0, +1, or +2 semitones. Notably, as will be discussed in Section~\ref{sec:main_result}, pitch shifting's inconsistent benefits across datasets were resolved by MoE models' increased capacity.

\section{Experiments}
\begin{table}[t]
\footnotesize
\begin{tabular}{
  >{\small\centering\arraybackslash}p{3.5cm}
  >{\small\centering\arraybackslash}p{3.5cm}
}
\toprule
 \it Train & \it Test \\
\midrule
 \hlgray{MusicNet-EM}, GuitarSet, \hlpink{MIR-ST500}, \hlpink{ENST-Drums}, \hlgray{Slakh}, EGMD, Maestro, CMedia, \hlgray{URMP}, SMT-Bass & \hlgray{MusicNet}, \hlgray{MusicNet-EM}, GuitarSet,\hlpink{MIR-ST500}, \hlpink{ENST-Drums}, \hlgray{Slakh}, Maestro, MAPS, \hlgray{URMP},  \hlgray{RWC-Pop (refined)}\\

\bottomrule
\end{tabular}
\caption[]{Summary of datasets for train/test. Multi-instrument datasets with full annotation and stems are highlighted in light blue, while those with partially annotated instruments are highlighted in pink. (refined) We offer updated annotations for RWC-Pop~\cite{rwc}. } 
\label{tab:train_eval_dataset}
\end{table}

\subsection{Experimental Setup}
\noindent\textbf{Data Preparation:} Table~\ref{tab:train_eval_dataset} lists the datasets used for training and evaluating our model. We offer a software package for dataset setup and split information to ensure reproducibility of our results. Audio data was converted into 16 kHz mono WAV format. Stems were stored as arrays, and mix-tracks as WAV files, also treating stemless tracks as mix-tracks. For training our \texttt{Single} decoder models on MIR-ST500~\cite{mirst500} and CMedia~\cite{mirex2020}
, we produced singing and accompaniment stems using a pre-trained separation model~\cite{spleeter}. With the \texttt{Multi} decoder, we also incorporated the original mix tracks from these datasets.

\noindent\textbf{Evaluation Metrics:} To evaluate transcription accuracy for each instrument, we employ the \textit{Instrument Note Onset F1} metric ~\cite{perceivertf}. This metric, valid for any instruments including drums, requires matching the onset, pitch, and instrument to the reference within a tolerance of ±50 ms. For multiple non-drum instruments, we additionally utilize the \textit{Instrument-Agnostic Onset F1 and Offset F1} necessitating exact matches for only onset or both onset and offset. These metrics parallel the standard \textit{Note F1} metrics~\cite{mir_eval} for single-instrument datasets. Furthermore, we used the \textit{Multi (instrument offset) F1} metric~\cite{mt3} for evaluating multi-instrument AMT systems, where correct predictions require matching onset-offset pairs, pitch, and instrument type, excluding drum offsets. Our Multi F1 metric is notably more stringent than the Multi Onset F1 reported for PerceiverTF~\cite{perceivertf}.

\noindent\textbf{Vocabulary:} Our models were trained using \texttt{MT3\_FULL\_PLUS} and tested on \texttt{MT3\_MIDI\_PLUS}, detailed in Section F of the Supplemental Document. Despite testing exclusively with the \texttt{MIDI} vocabulary, results in Table~\ref{tab:multi_result}, labeled \emph{+full vocab}, show that training with the more fine-grained \texttt{FULL} vocabulary enhanced performance compared to training and testing solely with \texttt{MIDI}.
 
\begin{table*}[ht]
\centering
\footnotesize
\begin{tabular}{
  >{\raggedright\arraybackslash}m{3cm}
  >{\raggedright\arraybackslash}m{1.1cm}
  >{\centering\arraybackslash}m{0.9cm}
  >{\centering\arraybackslash}m{1.5cm}
  >{\centering\arraybackslash}m{1.5cm}
  >{\centering\arraybackslash}m{1.5cm}
  >{\centering\arraybackslash}m{1.5cm}
  >{\centering\arraybackslash}m{1.1cm}
  >{\arraybackslash}m{1.5cm}
}
\toprule
\it Test Set & \it Instrument & \texttt{YMT3} & \texttt{YMT3+} & \texttt{YPTF+S} & \texttt{YPTF+M} & \texttt{YPTF.MoE+M} & \it MT3~\cite{mt3}  & \it AMT \\
\cmidrule(lr){3-7}
 & & \it noPS & \it noPS $|$ PS~~~~~ & \it noPS $|$ PS~~~~~ & \it noPS $|$ PS~~~~~ & \it noPS $|$ PS~~~~~ & (colab) & Baseline *\\
\midrule
MAPS~\cite{maps} (unseen) & \multirow{2}{*}{Piano}  & 81.44 & 85.92 $|$ 87.73 & 88.37 $|$ \textbf{88.73} & 87.84 $|$ 86.88 & 87.88 $|$ 86.25 & 80.62 & 88.40~\cite{edwards2024data}$ \clubsuit$ \\
MAPS~\cite{maps} (seen) & &-&-&-&-&-&-& 85.14~\cite{hftt}$ \clubsuit$ \\
Maestro v3 &  & 94.78 & 94.80 $|$ 94.31 & 96.28 $|$ 95.85 & 95.59 $|$ 94.54 & 96.98 $|$ 96.52  & 94.86 & \textbf{97.44}~\cite{hftt}$ \clubsuit$ \\
\midrule
\multirow{2}{2cm}{\parbox{2cm}{\footnotesize MusicNet ext.\newline (EM)~\cite{musicnet_em}}} 
 & Strings & 81.69 & 89.04 $|$ 88.34 & 88.39 $|$ 89.39 & 88.52 $|$ 87.04 & \textbf{91.32} $|$ 90.07 & -$\triangle$ & 80.00~\cite{musicnet_em} $\sharp$\\
 & Winds & 74.95 & 82.91 $|$ 80.53  & 77.72 $|$ 79.59 & 77.18 $|$ 76.54 & 83.46 $|$ 78.50 & -$\triangle$ & \textbf{85.50}~\cite{musicnet_em} $\sharp$\\
\midrule
\multirow{2}{2cm}{\parbox{2cm}{\footnotesize MusicNet ext.\newline ~\cite{reconvat, musicnet_em}}} 
 & Strings & 58.20 & 64.67 $|$ 63.94 & 64.63 $|$ 65.40 & 64.17 $|$ 64.08 & \textbf{66.14} $|$ 66.09 & -$\triangle$ & 63.90~\cite{musicnet_em} $\sharp$\\
 & Winds & 50.76 & 55.58 $|$ 55.05 & 52.55 $|$ 54.27 & 51.82 $|$ 51.42 & 55.95 $|$ 55.33  & -$\triangle$ & \textbf{60.90}~\cite{musicnet_em} $\sharp$\\
\midrule
\footnotesize MIR-ST500~\cite{mirst500} (SVS) & \multirow{3}{*}{Singing} & 67.98 & 70.39 $|$ 70.69  & 70.82 $|$ 70.56 & 71.07 $|$ 71.32 & 71.60 $|$ \textbf{72.05} & -$\lozenge$ & 70.73~\cite{avsvt} \\
\footnotesize MIR-ST500~\cite{mirst500} &  & ~~3.62 & 64.03 $|$ 65.69  & 66.75 $|$ 67.11 & 69.67 $|$ 70.26 & 70.59 $|$ 71.07 & -$\lozenge$  & \textbf{78.50}~\cite{perceivertf} \\
\footnotesize MIR-ST500 (100ms~\cite{mirex2020}) &  & ~~3.64 & 71.15 $|$ 72.08 & 73.26 $|$ 73.89 & 79.29$|$ 80.63 & 81.14$|$ \textbf{82.08} & -$\lozenge$ & - \\
\midrule
\footnotesize ENSTdrums (DTP~\cite{wu2018review}) & \multirow{2}{*}{Drums} & 87.77 & 87.60 $|$ 87.40 & 89.72 $|$ \textbf{90.65} & 88.68 $|$ 90.61 & 88.79 $|$ 89.48 & 77.82 & 84.50~\cite{wu2018review} $\clubsuit$\\
\footnotesize ENSTdrums (DTM~\cite{wu2018review}) &  & 78.64 & 81.84 $|$ 83.09 & 85.65 $|$ 86.41 & 85.14 $|$ 87.18 & 85.92 $|$ \textbf{87.27} & 70.31 & 79.00~\cite{wu2018review} $\clubsuit$ \\
\midrule
\footnotesize GuitarSet~\cite{guitarset} (MT3~\cite{mt3}) & Guitar & 88.53 & 91.39 $|$ 88.49 & 91.61 $|$ 88.32 & 88.92 $|$ 86.74 & \textbf{91.65} $|$ 88.87 & 89.10 & 91.10~\cite{perceivertf}\\
\midrule
\midrule
URMP~\cite{urmp} Onset F1~\cite{mt3} & Agnostic & 77.10 & 80.00 $|$ 81.47  & 81.11 $|$ 81.54 & 74.56 $|$ 75.72 &  81.05 $|$ \textbf{81.79}  & 76.65 & 77.0~\cite{mt3} \\
URMP~\cite{urmp} Multi F1~\cite{mt3} & Ensemble & 58.23 & 62.13 $|$ 62.03  & 64.34 $|$ 65.89 & 57.25 $|$ 59.82 &  67.22 $|$ \textbf{67.98}  & 58.71 & 59.0~\cite{mt3} \\
\midrule
Slakh~\cite{manilow2019cutting} Onset F1~\cite{mt3} & Agnostic & 64.83 & 77.96 $|$ 75.28 & 80.70 $|$ 76.32 & 79.39 $|$ 75.68 & 84.14 $|$ \textbf{84.56} & 75.20 & 81.9~\cite{perceivertf}\\
Slakh~\cite{manilow2019cutting} Multi F1~\cite{mt3} & All & 61.77 & 65.92 $|$ 63.61  & 69.52 $|$ 65.13 & 69.37 $|$ 64.96 &  73.98 $|$ \textbf{74.84}  & 57.69 & 62.0~\cite{mt3}$ \heartsuit$ \\
\bottomrule
\end{tabular}
\caption{Dataset-wise Note Onset F1. \textit{PS} and \textit{noPS} represent training with and without pitch shifting augmentation, respectively. (EM) denotes evaluation using refined labels~\cite{musicnet_em}. (SVS) refers to experiments using singing separated audio as input, obtained through \textit{Spleeter}~\cite{spleeter}. (DTP) represents using drum and percussion as input. (DTM) uses input including drum, percussion, and accompaniment. The Onset F1 score on Slakh is instrument-agnostic F1 for non-drum classes. ($\triangle$) Unavailable due to training split overlaps. ($\clubsuit$) Single-instrument AMT. ($\lozenge$) Singing voice class was not defined. ($\sharp$) Additionally collected synthetic data from 8.5K songs were used for pre-training~\cite{musicnet_em}.}
\label{tab:main_table}
\end{table*}

\noindent\textbf{Training:} Our models were trained with two NVIDIA A100 GPUs using BFloat16 mixed-precision. In the implemented online data pipeline, four CPU processes per GPU were allocated to efficiently load and augment data without causing streaming bottlenecks. In our preliminary experiments, we tested three optimizers at a constant learning rate of 1e-03: AdaFactor~\cite{shazeer2018adafactor}, AdamW~\cite{adamw}, and AdamWScale~\cite{adamwscale}. AdamWScale, a variant of AdamW that normalizes gradients using root-mean-square (RMS) energy, provided the most efficient training. Our models were trained using AdamWScale and a cosine scheduler for 300K steps, with initial and final learning rates of [1e-02, 1e-05] and a 1,000-step warm-up from 1e-03. We set the dropout rate at 0.05.

\begin{table}[ht]
\centering
\footnotesize
\begin{tabular}{
  >{\arraybackslash}p{2.5cm}
  >{\centering\arraybackslash}p{1cm}
  >{\centering\arraybackslash}p{1cm}
  >{\centering\arraybackslash}p{1cm}
}
\toprule
Model & Onset~F1 & Offset~F1 & Drum~F1 \\
\midrule
\texttt{YMT3} base & 64.8 & 41.7 & 77.8 \\
~~~\it{+ Intra-aug.} & +\bf 4.8 & +\bf5.5 & +0.6 \\
~~~\it{+ Full-vocab.} & +0.6 & +2.1 & +2.6\\
~~~\it{+ Data balancing} & + \bf4.0 & +\bf 4.7 & +1.3\\
~~~\it{+ Cross-aug.} & +\bf 4.0 & +\bf7.2 & +1.6\\
~~~\it{+ PTF-encoder} & +1.8 & +\bf4.2 & +1.9\\
~~~\it{+ FFN $\rightarrow$ MoE} & +1.5 & +1.3 & +3.7 \\
~~~\it{+ Multi decoder} & +1.8 & +\bf4.0 & +0.6 \\
\texttt{YPTF.MoE+Multi} & \textbf{84.6} & \textbf{70.7} & \textbf{90.1} \\
\midrule
MT3~(colab) & 75.2 & 56.8 & 83.9 \\
MT3~\cite{mt3} & 76 & 57 & - \\
PerceiverTF~\cite{perceivertf} & 81.9 & - & 78.3 \\
\bottomrule
\end{tabular}
\caption{Model component analysis and comparison on the Slakh~\cite{manilow2019cutting} dataset. (-) Values not reported.}
\label{tab:multi_result}
\end{table}

\begin{figure}[ht]
\centering
\includegraphics[width=0.79\columnwidth]{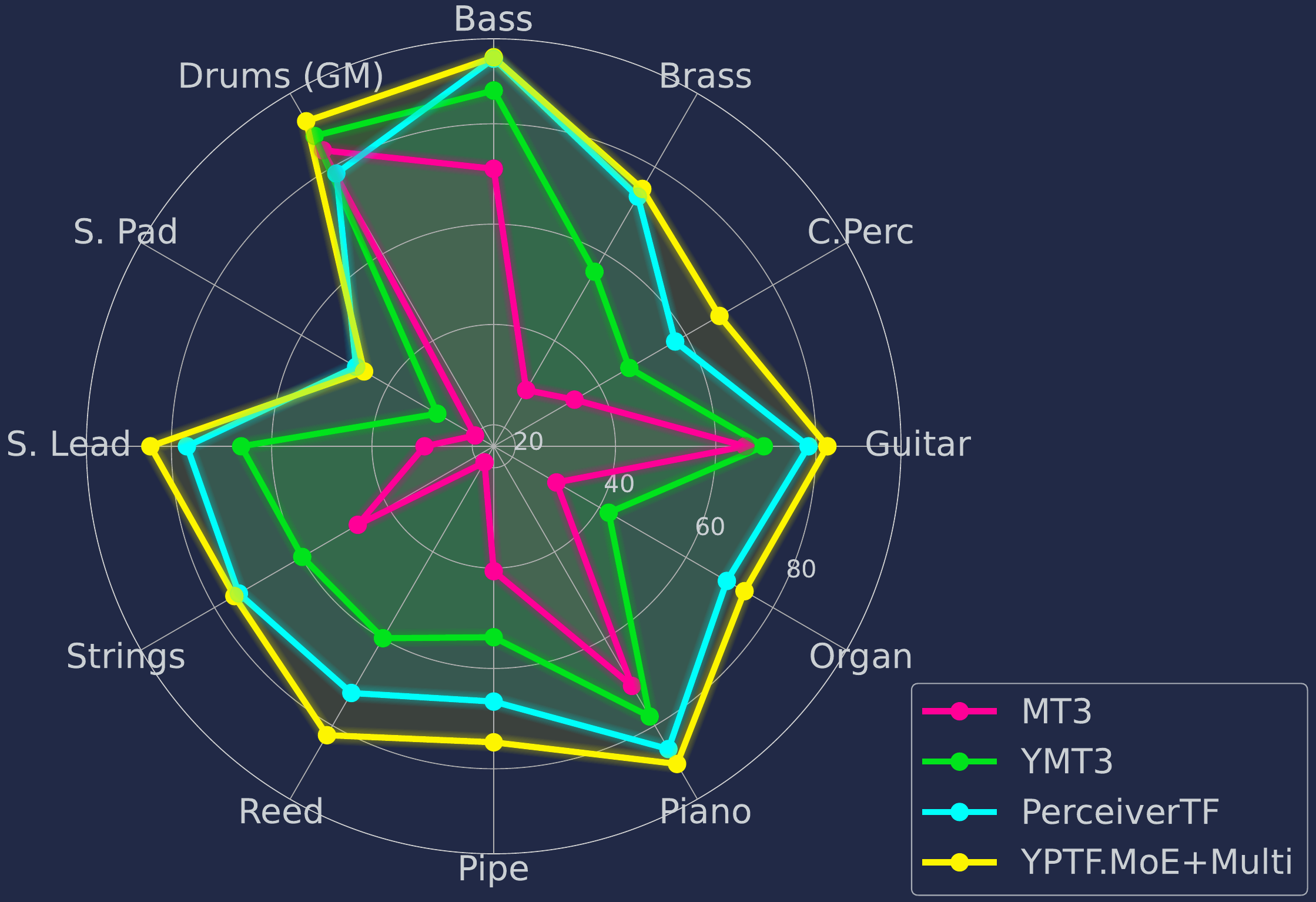} 
\caption{Instrument Onset F1 on Slakh~\cite{manilow2019cutting}. }  
\label{fig:skakh_radar}
\centering
\includegraphics[width=0.79\columnwidth]{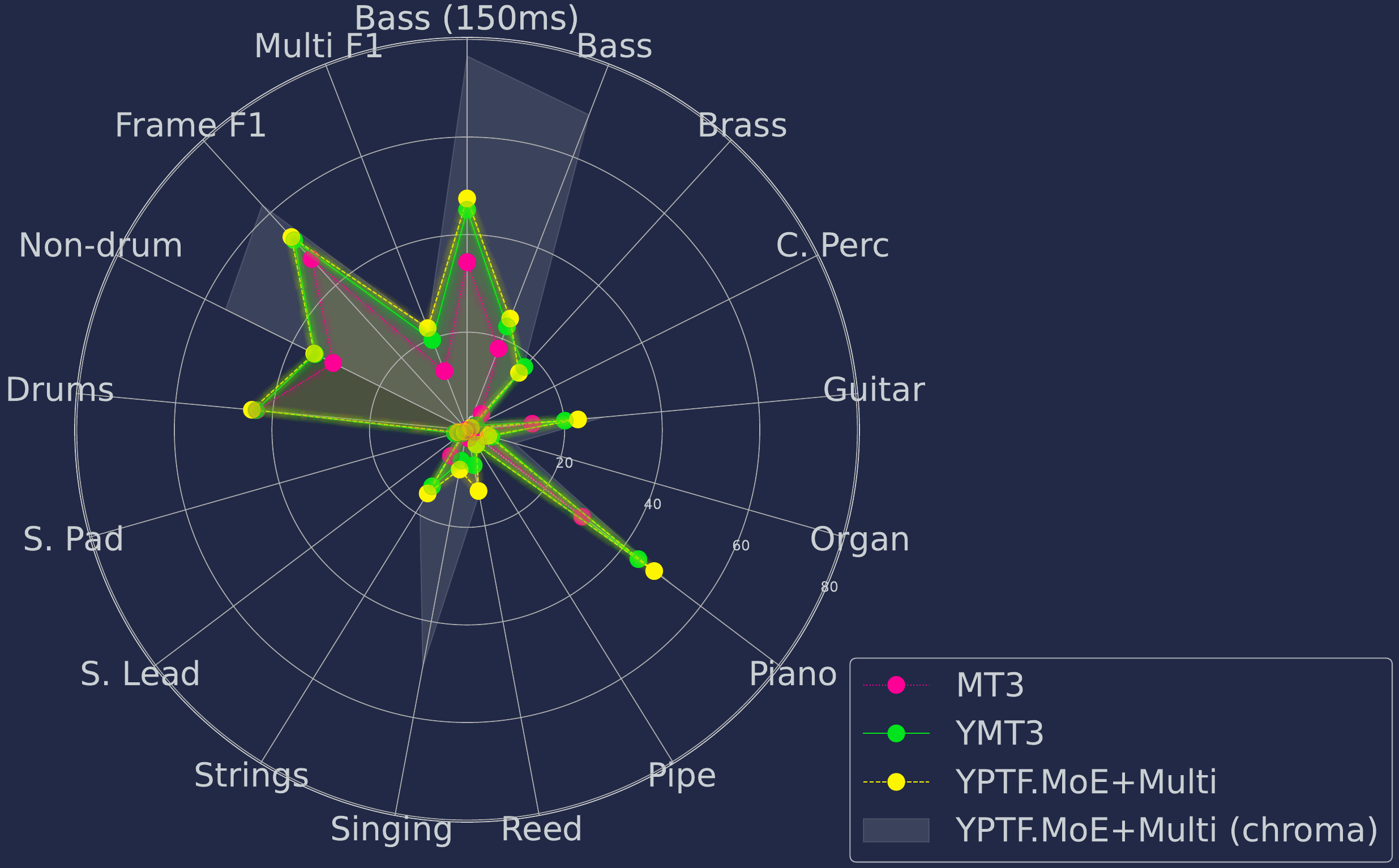} 
\caption{Instrument-Onset/Frame/Multi F1 on RWC-Pop~\cite{rwc}.}
\label{fig:rwc_radar}
\end{figure}

\subsection{Results and Discussion}
\label{sec:main_result}

In Table~\ref{tab:main_table}, our models are compared with other state-of-the-art models across datasets. From MAPS to GuitarSet, evaluations use Instrument Note Onset F1, while URMP and Slakh are assessed using Instrument-agnostic Note Onset F1 and Multi F1. Due to space constraints, only the top-performing baselines (*) are listed on the table's rightmost column. Details of all models are available in our project repository.

Our models prefixed by \texttt{Y-} outperformed MT3~\cite{mt3} across all datasets. Notably, our models and the unseen baseline~\cite{edwards2024data}, trained without MAPS~\cite{maps}, outperformed the baseline~\cite{hftt} trained on MAPS. This is likely due to the Maestro~\cite{maestro} dataset being about nine times larger, providing significantly more in-domain knowledge. Among our models, \texttt{YPTF.MoE+Multi} matched or exceeded the performance of the latest baseline models in most datasets. It showed exceptional performance on both refined and unrefined datasets in MusicNet strings, particularly in tests with refined labels (EM~\cite{musicnet_em}). However, a noticeable under-performance was observed in singing transcription compared to the baseline~\cite{perceivertf}. As evidenced by about 10\% higher F1 on the MIR-ST500 (100ms), many onset timing errors exceeded the acceptable 50ms range and fell within 100ms. Given that our model and the baseline~\cite{perceivertf} share similar encoder structures, our decoder may be more prone to timing errors than traditional piano-roll models. Additionally, the practicality of a 100ms onset tolerance, used in past MIREX~\cite{mirex2020}
singing transcription protocol, appears justified.

\texttt{YMT3+} and \texttt{YPTF+Single} differ only in their encoders. This comparison revealed that the PTF encoder architecture performs particularly well in complex multi-instrument datasets such as MIR-ST500, ENSTdrums (DTM), and Slakh. \textit{Cross-stem augmentation}, denoted by the (+) symbol in model names, proved essential for transcribing singing without singing voice separation (SVS). \texttt{YMT3} recorded an F1 score of 3.6\% without separation, while \texttt{YMT3+} with augmentation reached 64\%. The models with \texttt{Multi} decoders were beneficial when training on partially annotated datasets, such as MIR-ST500 and ENSTdrums. \textit{Mixture of Experts (MoE)} showed consistent performance improvements across all datasets. Notably, while pitch-shifting often led to performance degradation in other models, \texttt{YPTF.MoE} compensated for this loss or even improved performance, as evidenced by the Slakh result.

As compared in the lower section of Table~\ref{tab:main_table}, \texttt{YPTF.MoE+ Multi} significantly outperformed the baselines (MT3~\cite{mt3} and PerceiverTF~\cite{perceivertf}) on multi-instrument datasets such as URMP and Slakh. The baseline Multi F1 score marked with a $\heartsuit$ is from MT3 authors' report~\cite{mt3}. 
For the complete comparison table with MT3~\cite{mt3} and PerceiverTF~\cite{perceivertf}, see Section H of the Supplemental Document. 

\noindent\textbf{Ablation Study: }In Table~\ref{tab:multi_result}, the impact of each model component on performance was investigated. Both intra-~and cross-stem augmentations significantly improved performance by over 4 percentage points, while all other proposed components steadily enhanced transcription performance. Additionally, the performance improvement denoted by \textit{Data balancing} suggested that previously adopted temperature-based sampling in MT3~\cite{mt3} might not be suitable for determining the sampling probability of AMT datasets. This is further discussed in Section F of the Supplemental Document.

\noindent\textbf{Performance on Pop Music:} As seen at the bottom of Table~\ref{tab:multi_result}, our model demonstrated competitive performance on the synthetic dataset~\cite{manilow2019cutting} compared to other multi-AMT models. In Figure~\ref{fig:skakh_radar}, our final model achieved 50 to over 90\% performance for most instruments, except for a few non-mainstream ones like chromatic percussion (c. perc) and synth pad (s.pad) in the synthetic dataset. However, \textit{a significant limitation emerged in its performance on commercial pop music recordings}, as shown in Figure~\ref{fig:rwc_radar}. Particularly for non-main instruments (excluding piano, bass, vocals, and drums), our models performed below 10\%. This suggests potential biases introduced by training primarily on synthetic datasets, which may not fully cover the diverse timbres of pop music. Furthermore, except for the piano, all the pitched instruments showed a significant gap in the chroma-level metric, suggesting substantial octave errors and hinting that more varied pitch-shifting could be beneficial.

\section{Conclusion and Future Work}
This work presented \texttt{YourMT3+}, a hybrid model suite that combines MT3 and PerceiverTF features. Our final model, \texttt{YPTF.MoE +Multi}, employed spectral cross-attention and a Mixture of Experts in its encoder for enhanced performance, and a multi-channel decoder  to handle the instruments where annotation is partially available. Our models trained using the proposed online augmentation strategy demonstrated direct vocal transcription capabilities without the need for a singing separation front-end. The final model significantly outperformed MT3 and PerceiverTF on the multi-AMT benchmark with a parameter increase of less than 2.5\% compared to MT3. Evaluations across ten public datasets also validated our model's competitiveness. Despite progress, challenges persist: onset timing in singing voice transcription lags behind our baseline, and low performance in pop music may stem from reliance on synthetic datasets for diverse instruments. Future research will address these issues.

\bibliographystyle{IEEEbib-abbrev}
\bibliography{refs}


\end{document}


\ninept
\maketitle

\section{YouTube Transcription Demo}
\begin{figure}[h]
\centering
\includegraphics[width=\columnwidth]{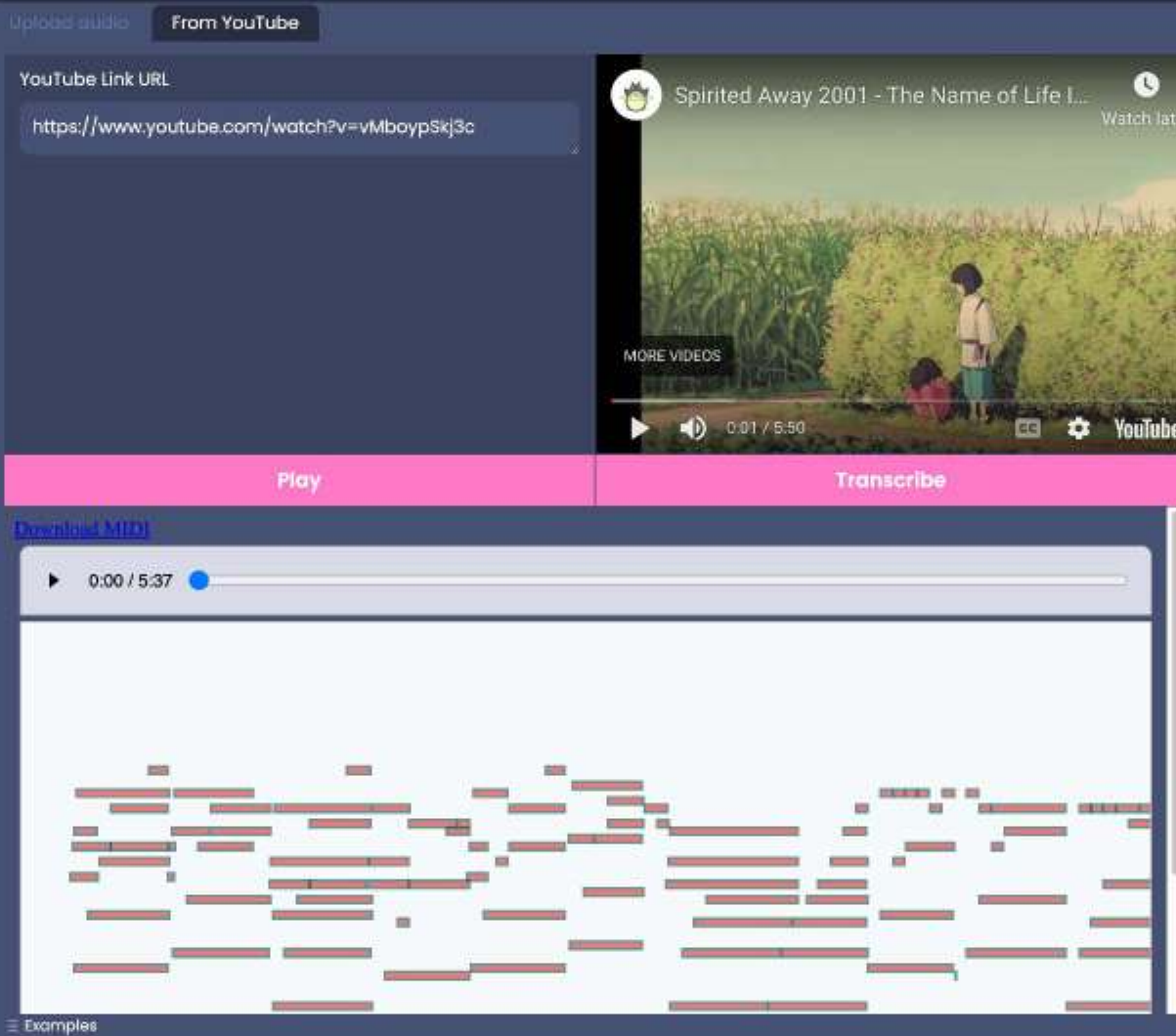} 
\caption{Grahpical user interface for YouTube Transcription.}
\label{fig_demo}
\end{figure}

For demonstration purposes, we provide a Colab notebook\footnote{\url{https://colab.research.google.com/drive/1AgOVEBfZknDkjmSRA7leoa81a2vrnhBG?usp=sharing}} that allows transcription from a YouTube link as in Figure~\ref{fig_demo}. In this demonstration experiment, using the NVIDIA T4 GPU provided for free on Google Colab in float32 operation mode, we were able to transcribe approximately six minutes of piano music within 40 seconds. In float16 operation mode, it took about 10 seconds on the T4 GPU, which is equivalent to 36$\times$ real-time. On GPUs from the \textit{Ampere} generation or later, which support 16-bit mixed precision, the process was completed in under 10 seconds.

One known issue was that when transcribing music from sources outside the dataset, models trained with Pitch-shift (PS) often incorrectly transcribed segments a semitone higher or lower. This issue was not observed in the models marked with (noPS), as tested in the example audio source\footnote{\url{https://youtu.be/9E82wwNc7r8?si=I-WyfwJXCBDY2reh}}.

\newpage

\section{Model details}

\subsection{Model Front-end Specification}

\begin{table}[ht]
\centering
\begin{tabular}{lll}
\toprule
\it Encoder Type & \texttt{YMT3} & \texttt{YPTF}\\
\midrule
length of $X$  & $2.048$ s & $2.048$ s\\
codec & \it mel-spec. & \it spec. \\
hop-size & $128$ ($8$ ms) & $300$ ($18.5$ ms)  \\
sample rate & $16,000$ & $16,000$ \\
input frames & $32,767$ & $32,767$ \\
n-FFT & $2048$ & $2048$ \\
n-Bin & $512$ & $1024$\\
pre-encoder & No & Yes\\
shape of $S$ & $256\times512$& $110\times128\times128$ \\
 & \small ($t\times f$) & \small($t\times c \times f'$) \\
\bottomrule
\end{tabular}
\caption{Input configuration parameters.}
\label{tab:audio_config}
\end{table}

\subsection{YMT3}
Our base model, \texttt{YMT3}, replicates MT3~\cite{mt3}, which is based on the \texttt{T5-small v1.1}\footnote{\url{https://github.com/google-research/text-to-text-transfer-transformer}} architecture. Sinusoidal positional encoding is added to the inputs for the encoder and decoder. 

\subsection{Pre-encoder Specification}
The model specification of pre-encoder for \texttt{YPTF} described in Section 3.1 is displayed in Table~\ref{tab:pre_enc_config}.

\begin{table}[ht]
\small
\centering
\begin{tabular}{ll}
\toprule
\it Parameter & \it Value \\
\midrule
Kernel Size & $(3, 3)$ \\
Average Pooling Kernel Size & $(1, 2)$ \\
Number of Conv Layers per Block & $2$ \\
Number of Blocks & $3$ \\
\bottomrule
\end{tabular}
\caption{PreEncoder specifications.}
\label{tab:pre_enc_config}
\end{table}

\newpage

\subsection{Number of Parameters}
\begin{table}[h]
\centering
\begin{tabular}{
  >{\small\arraybackslash}p{2cm}
  >{\centering\arraybackslash}p{1.5cm}
  >{\centering\arraybackslash}p{3.7cm}
}
\toprule
\it Model & \small\it \# layers Enc/Dec  & \small\it \# Parameters Total (Enc/Dec)  \\
\midrule
\texttt{YMT3}, \texttt{YMT3+} & 12/12& 44.7M (19.4M/25.7M) \\
\texttt{YPTF+Single} & 15/12 & 29.9M (1.8M/25.7M) \\
\texttt{YPTF+Multi} & 15/12 & 29.9M (1.8M/25.7M) \\
\texttt{YPTF.MoE+Multi} & 15/12 & 45.8M (20.3M/25.7M) \\
\bottomrule
\end{tabular}
\caption{Number of parameters. The total size is not a direct sum of the encoder and decoder due to additional components such as linear projection layers, LM-head, and pre-encoder layer.}
\label{tab:model_complexity}
\end{table}

\subsection{Mixture of Experts Layer in \texttt{YPTF.MoE}}
In Section 3.2, the Feed-Forward Networks (FFNs) within each latent transformer and temporal transformer of the PTF block are described as follows:
\begin{equation}
    \textrm{FFN}_\texttt{YPTF}(h_\textrm{att}) = \textrm{ReLU}(h_{\textrm{att}}\cdot W_1^T) \odot W_2,
\end{equation}
where \( h_{\textrm{att}} \) is the normalized output of the attention module, \( W_1 \) and \( W_2 \) are the weights of linear layers without bias terms, and \( \odot \) denotes an element-wise product. 

In the \texttt{YPTF.MoE} configuration, the standard FFN is replaced by a mixture of experts (MoE), in which each expert \( \epsilon \) operates as a Gated Linear Unit (GLU)~\cite{swiglu} activated by a Sigmoid Linear Unit (SiLU)~\cite{silu}:
\begin{equation}
    \epsilon(h_{\textrm{att}}) = (\textrm{SiLU}(h_{\textrm{att}}\cdot W_1^T) \odot (h_{\textrm{att}}\cdot V_{\textrm{gate}}^T)) \cdot W_2^T,
\end{equation}
where \( W_1 \), \( W_2 \), and \( V_{\textrm{gate}} \) are the weights of the linear layers. The MoE replacing FFN is further defined as:
\[
    \textrm{MoE}(h_{\textrm{att}}) = \sum^{n-1}_{i=0}\textrm{Softmax}(\textrm{Top2}(h\cdot W_g))_i \cdot \epsilon_i(h_{\textrm{att}}),
\]
where \( n = 8 \) represents the number of experts. The $\sum\textrm{Softmax}$ $(\textrm{Top2}(h\cdot W_g))$ function selectively routes to 2 out of 8 experts. For the latent transformer, the weight matrix \( W \) is defined as \( W \in \mathbb{R}^{B \cdot t \cdot k \times n} \), organizing the batch size \( B \), time steps \( t \), and number of latents \( k \) sequentially. For the temporal transformer, the weight matrix \( W_g \) is defined as \( W_g \in \mathbb{R}^{B \cdot k \cdot t \times n} \), altering the positions of time steps and latents.

In Table~\ref{tab:model_complexity}, the \texttt{YPTF.MoE} model has about 2.5\% more parameters than \texttt{YMT3} and about 18.5 M more than \texttt{YPTF}. However, since only 2 out of the 8 experts are activated during inference, the model complexity increases by just about 5\%. Initial experiments on the Slakh dataset with models utilizing Top1, Top2, and Top4 experts showed that the Top2 model had the best performance. The Top4 model experienced a performance drop of about 1 percentage point, while the Top1 model performed on par with \texttt{YPTF}.
\vspace{0.5cm}

\noindent\textbf{Limitations} In our model, unlike recent reports of performance improvements in decoder-only models~\cite{jiang2024mixtral} and fine-tuning~\cite{wu2023mole} with MoE, we have applied MoE only to the audio encoder while training from scratch. Our intention was to expand the encoder capacity to learn a wider range of audio representations. The main results in Table 2 and the ablation experiments in Table 3 suggest that MoE helps prevent performance degradation, especially when pitch shifting is applied during training. However, we have not yet identified clear patterns regarding which expert FFN is active on specific datasets or types of tokens, as well as the effect of weight initialization~\cite{li2024cumo}. This remains a topic for future research.

\subsection{Comparing Output Length of \texttt{Single} vs \texttt{Multi}}
The maximum sequence length for training, denoted as \(N\), varies depending on the type of decoder described in Section~3.4. The \texttt{Single} uses \(N_{\text{single}} = 1024\) as per MT3~\cite{mt3}, while the \texttt{Multi} uses a smaller \(N_{\text{multi}}\), due to its division into multiple program groups. \texttt{Multi} benefits from shorter sequences but faces memory constraints with its 13 output channels (\(k'=13\)). We set \(N_{\text{multi}}\) to 256 to keep token truncation loss under 0.015\%.

\subsection{Implementation of Masking Loss}
The masking loss for unannotated instruments mentioned in Section 3.4 is implemented by filling the target sequence of the respective channels (representing the unannotated program groups) with \texttt{PAD} tokens. During training loss calculation, the \texttt{PAD} tokens are generally excluded.

\newpage

\section{Token Definition}
Output tokens representing note events are summarized in Table~\ref{tab:token}. The structure of the note sequence follows the proposal in MT3~\cite{mt3}. It begins with the declaration of tie events using program and pitch tokens for notes continued from previous segments. Subsequently, the main events are organized chronologically, with simultaneous events sorted in the order of \{program, velocity, pitch\}. The sequence concludes with an EOS token followed by paddings. To avoid repeating tokens, \textit{run-length encoding}~\cite{robinson1967results} is employed.
\begin{table}[ht]
\small 
\centering
\begin{tabular}
{p{0.12\columnwidth}p{0.1\columnwidth}p{0.1\columnwidth}p{0.45\columnwidth}}
\toprule
\textit{Token Name} & \textit{Range} & \textit{Token Index} & \textit{Description} \\
\midrule
PAD & 0 & 0 & Special token for padding \\
EOS & 0 & 1 & Special token for end of sequence \\
UNK & 0 & 2 & Special token for unknown type event \\
shift & 0-205 & 3-\newline208 & Absolute time with `10 ms` grid within `2.048 sec` segments. \\
pitch & 0-127 & 209-336 & midi note numbers\\
velocity & 0-1 & 337-338 &\texttt{0} for note-on, \texttt{1} for off. \\
tie & 0 & 339 & A delimiter token declaring the end of annotating pre-activated notes. \\
program & 0-127 & 340-467 & \texttt{GM\_INSTR\_FULL} \cite{mt3} \\
drum & 0-127 & 468-595 & \texttt{GM\_DRUM} \cite{mt3}\\
\bottomrule
\end{tabular}
\caption{Token Definition}
\label{tab:token}
\end{table}
\newpage

\section{Data Augmentation}
\label{sup:data_augment}
\subsection{Default parameters for augmentation}
Default parameters for intra-stem augmentation and cross-dataset stem augmentation are outlined in Table~\ref{tab:cross_aug_param}. 

\begin{table}[h]
\centering
\begin{tabular}{ll}
\toprule
\textit{Parameter} & \textit{Default Value} \\ 
\midrule
$p$ & 0.7 \\
$\tau$ & 0.3 \\
$L$ & 1,024 \\
J & 5 \\
\bottomrule
\end{tabular}
\caption{Default parameters for intra-stem augmentation and cross-dataset stem augmentation.}
\label{tab:cross_aug_param}
\end{table}

\subsection{Stem Mixing Policy}
Stem mixing policy $\Psi$ with default values is outlined in Table~\ref{tab:stem_mix_policy}.

\begin{table}[h]
\begin{tabular}{ll}
\toprule
\textit{Parameter} & \textit{Value} \\ 
\midrule
Allowing instrument overlap & False \\
Mixing multiple drum tracks & False \\
Max number of subunit stems & 12 \\
$p_{\text{singing}}$, probability of retaining singing stem & 0.7 \\
\bottomrule
\end{tabular}
\caption{Stem mixing policy, $\Psi$ with default values.}
\label{tab:stem_mix_policy}
\end{table}
\newpage

\section{Dataset Curation}
\begin{itemize}
    \item MusicNet ext. \cite{thickstun}: This consists of 330 pieces of various classical music played as solo piano, piano trio, string trio, string quartet, winds quartet, ensemble, and so on. Since the original test split consists of only 3 pieces of music, we use the extended test split which consists of 10 pieces for evaluation. This split also has been widely used in recent works \cite{reconvat, musicnet_em}.
    \item MusicNet ext. (EM) \cite{musicnet_em}: This provides refined labels generated through an iterative re-alignment algorithm. Our models were trained using the refined labels\footnote{\url{https://github.com/benadar293/benadar293.github.io}} provided by the authors. The refined labels for the eight pieces were only missing in the training split: $2194, 2211, 2211, 2227, 2230, 2292, 2305$, and $2310$.
    \item GuitarSet \cite{guitarset}: This dataset comprises a total of 360 audio recordings performed by six guitarists. Each guitarist improvised three progressions in two versions across five musical styles and two tempi. Since there is no official test split, we created our own test split by randomly selecting four players per track for training, assigning one player for validation, and another player for evaluation. Given our split method exposed the model to all pieces multiple times, leading to slight over-fitting as observed in Riley et al.~\cite{riley_icassp2024}, future work should consider a piece-based data split.   
    \item MIR-ST500 \cite{mirst500}: The dataset comprises 500 pop songs in Chinese, English, and Korean, with 400 allocated for training and 100 for testing. All songs are licensed mix-tracks by professional artists. The provided annotations focus solely on singing and do not include annotations for the accompaniment. Additionally, we generated singing voice and accompaniment stems for the entire dataset, dubbed as \textit{MIR-ST500 (SVS)}. We used the 2-stem separation model from \texttt{Spleeter}\footnote{\spleeterurl} at 44.1 kHz sampling rate and then converted them to 16 kHz mono.
    \item ENST-Drums \cite{enstdrums}: For testing, subset of 21 files from "drummer 3," following the guidelines outlined in Tables 5 and 11 of the reference work \cite{wu2018review}. We utilise two distinct versions of audio files for testing: Drum Transcription in the presence of Percussion (DTP) and Drum Transcription in the presence of Melodic instruments (DTM). For DTM, the test set was mixed with a drums-to-accompaniment ratio of approximately -1.25 dB, as suggested in the survey~\cite{paulus2009drum}.
    \item RWC-Pop (bass) \cite{rwc}: This dataset was solely used for evaluating the bass transcription. Some MIDI files differed significantly from the actual transcriptions, and we speculate that they were primarily created as guide music before recording. While refining MIDI annotations, we encountered several alignment issues, and manually corrected them. We mainly focused on refining the bass tracks, setting aside other instruments. While correcting octave mistakes and pitch errors, we referenced the fundamental frequency in the spectrogram to maintain consistency. For doubled bass tracks, we separated them if there were distinguishable timbres; otherwise, we merged them. After reviews, we chose 90 out of 100 songs for the test set, excluding those without bass. More detailed information is in the dataset's work note file.
    \item Slakh \cite{manilow2019cutting}: The dataset consists of 2,100 multi-track MIDI pieces accompanied by professionally synthesised audio. We utilise the official train split for training on the full dataset training result. This dataset, although synthetic, is unique in its extensive coverage of audio stems and MIDI forms for instruments commonly found in pop music. This is a unique dataset that includes the bass instrument class along with a very small amount of SMT-Bass. However, we found that most of the 11 types of bass instruments included were transcribed one octave higher. We corrected this based on the F0 of the spectrogram. Meanwhile, the pre-trained MT3 model provided by the authors~\cite{mt3} used in benchmarks transcribes one octave higher, so we evaluated the model using uncorrected data for fair comparison.
    \item EGMD \cite{egmd}: This drum dataset has an official split information for train (35,218 files), validation (5,030 files), and test (5,290 files). This dataset includes approximately 433 hours of recorded performances by nine drummers, utilising 56 drum-kits included in the Roland TD-17. Despite its extensive dataset size, it also had the limitation of consisting mostly of repetitive drum performances.
    \item Maestro~\cite{maestro}: The dataset consists of approximately 200 hours of virtuosic piano performances, meticulously aligned with a fine precision of around 3 ms between note labels and audio wave-forms. We utilise the official split from version 3.0.0.
    \item MAPS \cite{maps}: The dataset comprises recordings of MIDI-aligned piano performances in classical music. We utilise only the test split from the splits used in the reference \cite{reconvat}.
    \item URMP~\cite{urmp}: The dataset contains 44 classical music pieces in different ensemble formats, with multi-stem audio and 10ms interval labels. Videos are available but not used. We follow the MT3 paper's split, with 35 training files and 9 testing files out of 44 total.
    \item CMedia: The dataset originates from the MIREX 2020 singing transcription task\footnote{\url{https://www.music-ir.org/mirex/wiki/2020:Singing_Transcription_from_Polyphonic_Music}} and consists of 100 songs from YouTube. It provides annotations for intervals and pitch, with a 100ms onset tolerance. We obtained this dataset directly from the author, and successfully corrected offset errors in six songs, which were approved by the dataset's author. We have utilised this dataset only for training.
    \item SMT-Bass~\cite{smtbass}: The dataset contains genuine bass recordings, distinguishing it from synthetic data in the Slakh~\cite{manilow2019cutting}. This dataset encompasses various bass guitar techniques and involves recordings of the chromatic scale using three different bass guitars, each with three distinct pickup settings. We selectively extracted playing styles from the \textit{plucking style} category and mapped them to MIDI programs 33-37. The original dataset files had a duration of 3-4 seconds, and we extended them to 7-8 seconds by adding 1.8 seconds of silence at both ends. Pitch annotation issues were identified, particularly concerning string and fret annotations. To improve the dataset, we used amp envelope-based on/off detection to select regions and verified pitch accuracy by comparing it with \texttt{Crepe}~\cite{crepe} algorithm predictions. As a result, 1327 out of 2332 files were reliable, split into 1061 for training and 266 for validation using an 8:2 ratio. The SMT-bass dataset is valuable for genuine bass recordings, addressing pitch issues, and improving bass performance in the \textit{RWC-Pop (bass)} dataset context.
    \item RWC-Pop (full): This dataset contains 100 pieces of Japanese or English pop music. However, it does not always align with General MIDI program numbers and may not represent actual transcriptions, making it less reliable than Geerdes. While we have thoroughly revised the bass parts, we lacked the time to do so for other instruments. Instead, we mapped the instruments in Table~\ref{tab:mt3_midi_ext_plus} based on keyword filtering of the given MIDI file track names.
\end{itemize}

\section{Vocabulary}
\label{sup:vocab}

\begin{center}
\small
\tablefirsthead{\toprule \textit{Instrument Name} & \textit{ID} \\ \midrule}
\tabletail{\multicolumn{2}{r}{(continued)} \\}
\tablelasttail{\bottomrule}
\bottomcaption{\texttt{MT3\_FULL\_PLUS} vocabulary as extended from the \texttt{FULL} vocabulary from the MT3~\cite{mt3}. An asterisk (*) represents internal IDs that are not tokenised.}
\label{tab:mt3_full_plus}
\begin{supertabular}{ll}
Acoustic Piano & 0, 1, 3, 6, 7 \\
Electric Piano & 2, 4, 5 \\
Chromatic Percussion & 8, 9, 10, 11, 12, 13, 14, 15 \\
Organ & 16, 17, 18, 19, 20, 21, 22, 23 \\
Acoustic Guitar & 24, 25 \\
Clean Electric Guitar & 26, 27, 28 \\
Distorted Electric Guitar & 29, 30, 31 \\
Acoustic Bass & 32, 35 \\
Electric Bass & 33, 34, 36, 37, 38, 39 \\
Violin & 40 \\
Viola & 41 \\
Cello & 42 \\
Contrabass & 43 \\
Orchestral Harp & 46 \\
Timpani & 47 \\
String Ensemble & 48, 49, 44, 45 \\
Synth Strings & 50, 51 \\
Choir and Voice & 52, 53, 54 \\
Orchestra Hit & 55 \\
Trumpet & 56, 59 \\
Trombone & 57 \\
Tuba & 58 \\
French Horn & 60 \\
Brass Section & 61, 62, 63 \\
Soprano/Alto Sax & 64, 65 \\
Tenor Sax & 66 \\
Baritone Sax & 67 \\
Oboe & 68 \\
English Horn & 69 \\
Bassoon & 70 \\
Clarinet & 71 \\
Pipe & 72, 73, 74, 75, 76, 77, 78, 79 \\
Synth Lead & 80, 81, 82, 83, 84, 85, 86, 87 \\
Synth Pad & 88, 89, 90, 91, 92, 93, 94, 95 \\
Singing Voice (Main melody) & 100 \\
Singing Voice (Chorus) & 101 \\
\midrule
Drums* & 128 \\
Unannotated* & 129 \\
\end{supertabular}
\end{center}

\begin{center}
\small
\tablefirsthead{\toprule \textit{Instrument Name} & \textit{ID} \\ \midrule}
\tabletail{\multicolumn{2}{r}{(continued)} \\}
\tablelasttail{\bottomrule}
\bottomcaption{\texttt{MT3\_MIDI\_EXT\_PLUS} vocabulary used for generating 16-channel-General MIDI files in our demonstration.}
\label{tab:mt3_midi_ext_plus}
\begin{supertabular}{ll}
Piano (acoustic) & 0, 1, 2, 3, 4, 5, 6, 7 \\
Chromatic Percussion & 8, 9, 10, 11, 12, 13, 14, 15 \\
Organ & 16, 17, 18, 19, 20, 21, 22, 23 \\
Guitar (clean) & 24, 25, 26, 27 \\
Guitar (distortion) & 28, 29, 30, 31 \\
Bass & 32, 33, 34, 35, 36, 37, 38, 39 \\
Strings + Ensemble & 40, 41, 42, 43, 44, 45, 46, 47, \\
 & 48, 49, 50, 51, 52, 53, 54, 55 \\
Brass & 56, 57, 58, 59, 60, 61, 62, 63 \\
Reed & 64, 65, 66, 67, 68, 69, 70, 71 \\
Pipe & 72, 73, 74, 75, 76, 77, 78, 79 \\
Synth Lead & 80, 81, 82, 83, 84, 85, 86, 87 \\
Synth Pad & 88, 89, 90, 91, 92, 93, 94, 95 \\
Singing Voice & 100, 101 \\
\end{supertabular}
\end{center}

\begin{center}
\small
\tablefirsthead{\toprule \textit{Instrument Name} & \textit{ID} \\ \midrule}
\tabletail{\multicolumn{2}{r}{(continued)} \\}
\tablelasttail{\bottomrule}
\bottomcaption{\texttt{MT3\_MIDI\_PLUS} vocabulary as extended from the \texttt{MIDI} vocabulary from the MT3~\cite{mt3}. This vocabulary has been used for evaluation comparing with previous works~\cite{mt3, perceivertf}.}
\label{tab:mt3_midi_plus}
\begin{supertabular}{ll}
Piano & 0, 1, 2, 3, 4, 5, 6, 7 \\
Chromatic Percussion & 8, 9, 10, 11, 12, 13, 14, 15 \\
Organ & 16, 17, 18, 19, 20, 21, 22, 23 \\
Guitar & 24, 25, 26, 27, 28, 29, 30, 31 \\
Bass & 32, 33, 34, 35, 36, 37, 38, 39 \\
Strings + Ensemble & 40, 41, 42, 43, 44, 45, 46, 47, \\
 & 48, 49, 50, 51, 52, 53, 54, 55 \\
Brass & 56, 57, 58, 59, 60, 61, 62, 63 \\
Reed & 64, 65, 66, 67, 68, 69, 70, 71 \\
Pipe & 72, 73, 74, 75, 76, 77, 78, 79 \\
Synth Lead & 80, 81, 82, 83, 84, 85, 86, 87 \\
Synth Pad & 88, 89, 90, 91, 92, 93, 94, 95 \\
Singing Voice & 100, 101 \\
\end{supertabular}
\end{center}


\clearpage

\section{Re-balancing Multi-dataset Sampling}
In multi-task learning, especially within NLP study~\cite{mbert, xlme, mt5, xlmr} for multilingual models, handling data size imbalances across languages has been one of the key issues. The sampling distribution for each language can be defined as \(\Theta = \{\theta(l)\}\), where \(l\) is the \(l\)-th language among a total of \(L\). The widely adopted approach, temperature-based sampling, calculates the sampling probability:\[
\theta(l) = \left( \frac{n(l)}{n_{\text{total}}} \right)^{\frac{1}{c}},
\]
 where \(n(\cdot)\) denotes dataset size and the temperature $c$ is typically set $1.43$ (~\cite{mbert, xlme}) or $3.33$ (in~\cite{mt5, xlmr}, and also in MT3~\cite{mt3}). As \(c \rightarrow \infty\), \(\Theta\) tends towards uniformity\footnote{In multi-lingual large model training, extremely balanced sampling often resulted in biased models due to the over-repetition of smaller datasets, as revealed in previous study~\cite{unimax}.}. 

We experimented with various $c$ values in temperature-based sampling, but this did not prevent performance degradation of up to 5\% across multiple datasets. Several factors might contribute to this phenomenon:
\begin{itemize}
\item Data size calculation: NLP used tokenised lengths, while MT3 counted files, possibly oversimplifying complex datasets with varied audio lengths.
\item Repetitive data impact: Datasets like EGMD contain lengthy, repetitive loops. Data size becomes an unreliable metric for calculating $\Theta$ in such cases.
\item Dynamic data addition: Unlike MT3, our setup with data augmentation and continuous dataset integration necessitates constant updates to the optimal temperature setting.
\end{itemize}

To delve into this matter, we adopted a straightforward iterative method with \(\eta=5\) cycles to refine the dataset sampling weights \(\Theta_\eta\) for various datasets. The manual tuning followed this procedure:
\begin{enumerate}
    \item Initial weights \(\Theta_0\) were set using the temperature-based approach suggested by MT3~\cite{mt3}.
    \item We identified the dataset most prone to over-fitting, as shown by its validation loss curve.
    \item We redistributed 10\% of the weight from the over-fitting dataset evenly among the others, and train a new model from scratch.
    \item We repeated step 2 and 3, \(\eta\) times.
\end{enumerate}
The method's primary drawback was extending the total training time by \(\eta\) times and the challenge of clearly pinpointing over-fitting in step 2. We repeated this process about seven times to find a relatively good dataset balance. This value is shown in Table~\ref{tab:sample_prob}.

\begin{table}[ht]
\small 
\centering
\begin{tabular}
{p{0.5\columnwidth}p{0.4\columnwidth}}
\toprule
\textit{Dataset} & \textit{Sample Prob.} \\
\midrule
Slakh & 0.295\\
MusicNet (em)& 0.19\\
MIR-ST500 & 0.191\\
ENSTdrums & 0.05\\
GuitarSet & 0.01\\
EGMD & 0.004\\ 
URMP & 0.1\\ 
Maestro & 0.1\\ 
SMT Bass & 0.01\\ 
CMedia & 0.05\\
\bottomrule
\end{tabular}
\caption{Re-balanced Dataset Sampling Probability}
\label{tab:sample_prob}
\end{table}

We believe these experimental results not only highlight the limitations of temperature-based dataset sampling but also offer valuable insights into the importance of dataset re-balancing. Given these lessons, we plan to explore online meta-learning techniques~\cite{chen2024online} for an episode-based optimizations of \(\Theta\) in future work. 

\clearpage
\section{Performance Benchmark on Slakh}

In Table~\ref{tab:result_slakh_all}, we compare our final model, \texttt{YPTF.MoE+Multi}, with MT3~\cite{mt3} and PerceiverTF~\cite{perceivertf} using various metrics. The table displays three versions of MT3s. The first from the left is the result tested with our implemented metric on the Colab notebook provided by the MT3 author, the second is the result reported by the PerceiverTF author, and the third is the result reported by the MT3 authors.

At the top of the table, our model significantly outperformed other models in the Instrument Note Onset F1 metric for 11 instrument classes, falling slightly behind only the PerceiverTF in the "synth-pad" class.

In the middle section of the table, our model also led in the Instrument agnostic metrics.
\begin{table*}[hb]
\centering
\small
\begin{tabular}{
  >{\raggedright\arraybackslash}p{4cm}
  >{\centering\arraybackslash}p{1.8cm}
  >{\centering\arraybackslash}p{1.8cm}
  >{\centering\arraybackslash}p{1.8cm}
  >{\centering\arraybackslash}p{1.1cm}
  >{\centering\arraybackslash}p{1.8cm}
}
\toprule
\it Metric & \texttt{YPTF.MoE+M} (PS) & \it MT3~\footnote{https://github.com/magenta/mt3} \newline (our colab) & \it MT3 (colab~\cite{perceivertf}) & MT3~\cite{mt3} & PerceiverTF \cite{perceivertf} \\
\midrule
Bass & \textbf{93.20} & 71.03 & 90.6 & - & 93.0\\
Brass & \textbf{74.96} & 28.67 & 43.3& - & 73.2\\
C.Perc & \textbf{67.70} & 34.31 &34.31 &- & 57.5\\
Guitar & \textbf{82.27} & 65.90 & 73.2& -& 78.5\\
Organ & \textbf{73.48} & 30.14 & 36.3& -& 69.4\\
Piano & \textbf{88.84} & 70.87 & 78.0& -& 85.4\\
Pipe & \textbf{74.72} & 40.60 & 28.2& -& 66.6\\
Reed & \textbf{82.22} & 19.41 & 44.0& -& 72.5\\
Strings and Ens. & \textbf{75.44} & 47.02 & 55.1& -& 74.4\\
Synth Lead & \textbf{84.19} & 29.51 & 40.9&- & 76.9\\
Synth Pad & 45.57& 20.02 & 23.4& -& \textbf{47.4} \\
Drums & \textbf{90.05} & 83.85 & 77.3& -& 78.5\\
\midrule
Onset F1 (non-drum)  &\textbf{84.56} &75.20 & -    & 76 & 81.9\\
Offset F1 (non-drum) &\textbf{70.70} &56.78 & -    & 57 & - \\
Onset F1 (drums/gm)  &\textbf{90.05} &83.85 & 77.3 & -  & 78.5 \\
\midrule
Multi (Onset-Offset) F1 \cite{mt3} $\spadesuit$ & \textbf{74.84} & 57.16 & - & 62 & -\\  
Multi Onset F1 \cite{perceivertf} & - & - & 74.3 & - & 79.8 \\ 
\bottomrule
\end{tabular}
\caption{Performance metrics on Slakh~\cite{manilow2019cutting} dataset. The highest and second-highest scores are highlighted in boldface. $\spadesuit$: We use \texttt{MT3\_MIDI\_PLUS} vocabulary, which is equivalent to \texttt{MIDI}~\cite{mt3} plus singing. ($-$) Values not reported.}
\label{tab:result_slakh_all}
\vspace{6cm}
\end{table*}

The bottom section displays metrics comparing the overall performance of Multi-instrument AMT. For a fair comparison, we do not directly compare our Multi (Onset-Offset) F1 with the PerceiverTF authors' Multi Onset F1~\cite{perceivertf} score, which counts a note as correct if its program and onset match the reference. Our Multi F1 metric, following prior work~\cite{mt3}, counts a note as correct only if both its program, onset, and offset match the reference, thus indicating a more rigorous and comprehensive performance assessment.
\clearpage

\section{YourMT3 Toolkit}
\begin{figure}[h]
\centering
\includegraphics[width=\columnwidth]{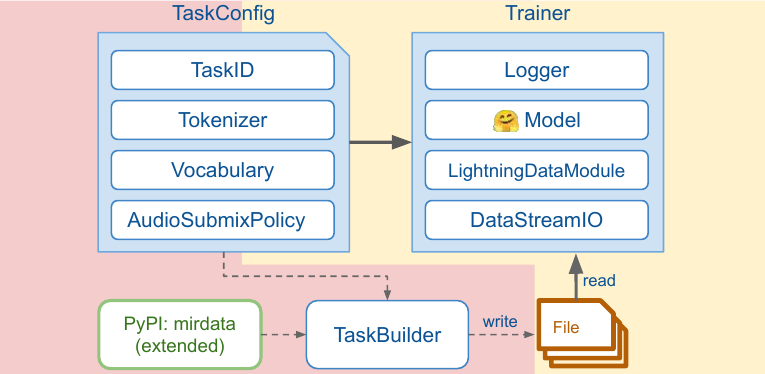} 
\caption{An overview of how to prepare a single MIR task (pink block), and train (yellow block) a model on the task. The dotted line represents preprocessing that runs only once during the task build, and the solid line represents streaming of data during training.}
\label{fig_yourmt3_toolkit}
\end{figure}

\begin{table}[ht]
\centering
\small
\begin{tabular}{p{0.45\columnwidth}p{0.45\columnwidth}}
\toprule
\textit{MT3} \cite{mt3} & \textit{YourMT3 \& YourMT3+} \\
\midrule
\textit{Disk I/O} & \\
$\bullet$ \texttt{TF-Record}, serialised data with fixed data shape & $\bullet$ Custom stem file format + 
FIFO cache for efficient random audio sampling\\
\midrule
\textit{Note Processing Pipeline} & \\
\addlinespace
MIDI $\leftrightarrow$ Notes $\leftrightarrow$ NoteSeq $\leftrightarrow$ Events $\leftrightarrow$ \textbf{Tokens($\heartsuit$)}
& MIDI $\leftrightarrow$ Notes $\leftrightarrow$ \textbf{Note-Events($\heartsuit$)} $\leftrightarrow$ Events $\leftrightarrow$ Tokens \\
\midrule
\textit{Stem Augmentation} & \\
$\bullet$ A few versions of offline processed data \par
$\bullet$ Dropping random stems & $\bullet$ Infinite random processing \textit{on-the-fly} \\
& (\textit{YourMT3+}) \par$\bullet$  Intra/Cross stem augmentation and pitch-shifting \\
\midrule
\textit{Optimizer} & \\
$\bullet$  AdaFactor \cite{shazeer2018adafactor} with constant learning rate & $\bullet$ AdamWScale \cite{adamwscale} with cosine scheduler\\
\midrule
\textit{Datasets for Training} &\\
$\bullet$ Collection of 6 datasets \cite{mt3}: Slakh, Cerberus, MusicNet, GuitarSet, URMP, Maestro & $\bullet$ Collection of 6 datasets: Slakh, EGMD, MusicNet EM, GuitarSet, ENST-Drums, 
MIR-ST500 + \textit{Spleeter} \\
\\
& (\textit{YourMT3+})
\par$\bullet$ Additional 4 datasets: Maestro, URMP, CMedia + \textit{Spleeter}, IDMT-SMT Bass 
\\

\bottomrule
\end{tabular}
\caption{Comparison of our implemented data pipeline and augmentation method with previous work~\cite{mt3}.}
\label{tab:comparison}
\end{table}
We developed a toolkit titled \texttt{YourMT3}~~\cite{chang2022yourmt3}
for training AMT models. The toolkit in Figure~\ref{fig_yourmt3_toolkit} comprises two primary components: the task and the trainer. Below are several design considerations aimed at streamlining training for multi-task learning with both audio and symbolic music data.

\begin{itemize}
    \item \textbf{Defining AMT tasks}: A task is simply definable with a set of MIDI tokeniser, vocabulary, and an audio processor. Vocabulary interacts with tokeniser, and  together with audio processor it configures the data-stream for mixing sub-tracks.

    \item \textbf{Processing on-the-fly}: The token storage in MT3~\cite{mt3} is sub-optimal for on-the-fly augmentation due to the absence of links between note onsets and offsets. To efficiently extract and blend short segments from large tracks at random start points, we introduced a self-contained data class \textit{NoteEvent}  and summarized in Table~\ref{tab:comparison} with $\heartsuit$. This approach satisfies the requirements for data processing speed and further adaptability to various data augmentations.
    
    \item \textbf{Dependency-free}: While MT3 processes notes through \texttt{NoteSeq}\footnote{\url{https://github.com/magenta/note-seq}}, an undocumented library rooted in \texttt{Pretty MIDI}\footnote{\url{https://github.com/craffel/pretty-midi}} leading to timing complications with overlapping notes, we constructed the pipeline utilizing the foundational \texttt{MIDO}\footnote{\url{https://github.com/mido/mido}} library.

\end{itemize}

Previous works~\cite{mt3, perceivertf} have adopted offline processing for data augmentation. In \texttt{YourMT3}, we implemented a pipeline for efficiently streaming thousands of audio stem files. This was achieved through a caching method and the definition of a custom note event format to rapidly process and tokenize note events. We have further enhanced our toolkit, now called \texttt{YourMT3+}. The advancements are highlighted in Table~\ref{tab:comparison}. 
This toolkit update provides an AMT training environment with minimized bottlenecks in distributed data-parallel (DDP) settings.

\clearpage
\bibliographystyle{IEEEbib}
\bibliography{refs}